# Three-dimensional realizations of flood flow in large-scale rivers using the neural fuzzy-based machine-learning algorithm

Zexia Zhang[1], Ajay B. Limaye[2], and Ali Khosronejad[1*]

[1]Civil Engineering Department, Stony Brook University, Stony Brook, NY 11794, USA

[2]Department of Environmental Sciences, University of Virginia, Charlottesville, VA 22904, USA

*Corresponding author, Email: ali.khosronejad@stonybrook.edu

**Abstract**

Machine learning methods have been extensively used to study the dynamics of complex fluid flows. One such algorithm, known as adaptive neural fuzzy inference system (ANFIS), can generate data-driven predictions for flow fields but has not been applied to natural geophysical flows in large-scale rivers. Herein, we demonstrate the potential of ANFIS to produce three-dimensional (3D) realizations of the instantaneous flood flow field in several large-scale, virtual meandering rivers. The 3D dynamics of flood flow in large-scale rivers were obtained using large-eddy simulation (LES). The LES results, i.e., the 3D velocity components, were employed to train the learnable coefficients of an ANFIS. The trained ANFIS, along with a few time-steps of LES results (precursor data) were then used to produce 3D realizations of flood flow fields in large-scale rivers with geometries other than the one the ANFIS was trained with. We also used the trained ANFIS to generate 3D realizations of river flow at a discharge other than that the ANFIS was trained with. The flow field results obtained from ANFIS were validated using separate LES runs to assess the accuracy of the 3D instantaneous realizations of the machine learning algorithm. An error analysis was conducted to quantify the discrepancies among the ANFIS and LES results for various flood flow predictions in large-scale rivers.

# 1 Introduction

Recent advances in high-performance computing have enabled researchers to carry out high-fidelity flow simulation of large-scale rivers using the computational fluid dynamics (CFD) method [1–8]. Yet due to the complex bathymetry, large scale, and high Reynolds number for flood flows in natural rivers, such high-fidelity simulations are expensive and require extensive computational resources. As a result, data-driven machine-learning methods have been extensively applied to study the fluid dynamics of complex flows [9–18]. For example, Hanna et al. [19] used an artificial neural network (ANN) and random forest regression (RFR) to predict the computational error associated with low-resolution CFD. Bakhtiari and Ghassemi [16] used feedforward neural networks (FNN) to predict hydrodynamic coefficients of a propeller as a function of blade number and the ratio of blade thickness to Marine cycloidal propeller diameter. Ti et al. [20] used ANN to correlate wake characteristics of wind turbines with the inflow parameters. Duraisamy et al. [21] studied the potential of machine-learning to improve the accuracy of closure models for turbulent and transition flows. Tracey et al. [22] demonstrated the potential of machine-learning algorithms to enhance and/or replace the traditional turbulence models such as Spalart-Allmaras. Ling et al. [23] employed a novel ANN architecture, i.e. the tensor basis neural network, to improve the accuracy and performance of Reynolds-averaged Navier Stokes (RANS)-based turbulence modelling. Singh et al. [24] used neural network to enhance the Spalart–Allmaras model in a RANS solver to predict strong adverse pressure-gradient flow over airfoils. Zhu et al. [25] constructed a new turbulent model using radial-basis-function neural network in order to directly map the turbulent eddy viscosity using the mean flow variables at high Reynolds number. Using the gene-expression programming algorithm, Zhao et al. [26] developed an explicit Reynolds-stress model directly implemented into RANS equations.

Data-driven machine-learning algorithms have also been applied to develop reduced-order models (ROM) for flow field prediction. For example, Mohan and Gaitonde [27] employed proper orthogonal decomposition (POD) and Long short-term memory (LSTM) architecture to develop a ROM for turbulent flow control. Lui and Wolf [28] developed a flow field predictive method using deep feedforward neural network (DNN). Data-driven methods have also been employed to produce super-resolution realizations. For example, Deng et al. [29] utilized a super-resolution generative adversarial network (SRGAN) and enhanced-SRGAN (ESRGAN) to augment the spatial resolution of measured turbulent flows. Using a convolutional neural network (CNN) and hybrid downsampled skip-connection/multi-scale (DSC/MS) model, Fukami et al. [15, 16] generated high-resolution flow fields from coarse flow field data. They also examined the performance of four machine learning methods (i.e., multilayer perceptron, random forest, support vector regression, and extreme learning machine) in a number of regression problems for fluid flows [31]. Liu et al. [32] developed the static convolutional neural network (SCNN) and the novel multiple temporal paths convolutional neural network (MTPC) to conduct super-resolution reconstruction of turbulent flows from direct numerical simulation (DNS).

The adaptive neural fuzzy inference system (ANFIS) [33] is a combination of fuzzy logic and ANN. It has been widely applied in hydraulics and fluid mechanics for several regression-based problems. For example, Gholami et al. [34] employed ANFIS to predict axial velocity and flow depth in a 90° open-channel bend using the flow discharge as an input. Qasem et al. [35] optimized ANFIS to predict the minimum flow velocity to prevent sediment deposition in open-channel flows. Şamandar [36] and Moharana and Khatua [37] used ANFIS to predict Manning's roughness coefficient of channels for uniform open-channel flows. ANFIS has been widely applied to predict flow discharges of rivers under base and flood flow conditions [23-34]. For example, Pasaie et al. [38] used ANFIS to predict the flow discharge in compound open channels. Rezaeianzadeh et al. [39] compared the accuracy of ANN, ANFIS, and two regression models in forecasting the maximum daily flood flow based on precipitation and shows that ANFIS's predictions are relatively more accurate than the other methods. He et al. [40] examined the ability of ANFIS, ANN, and support vector machine (SVM) to predict the flow rate in rivers based on antecedent data, and their analysis showed similar performance for the examined methods. Firat and Turan [41] compared ANFIS, feed forward neural networks (FFNN), and autoregressive (AR) methods in forecasting monthly flow rate of rivers and found that the performance of ANFIS is better than for the other methods. Overall, ANFIS is reported to be an efficient machine-learning algorithm for predicting intricate nonlinear regression between an input and output signals.

Despite the extensive assessments of ANFIS for regression-based predictions, no prior study has focused on ANFIS as a machine-learning algorithm to generate 3D realizations of large-scale river flow field under flood conditions. In this study, we employ the ANFIS machine-learning algorithm to predict the 3D velocity field of large-scale rivers under flood conditions in which flow fully fills the channel. The large-eddy simulation (LES) method is used to produce the training dataset required to train the machine-learning algorithm. Additionally, separately conducted LES will be used to evaluate the performance of the ANFIS in predicting the 3D velocity field of the rivers. The LES results of the model were also validated using a series of experimental data. The capability of the developed ANFIS machine-learning algorithm was examined at different Reynolds numbers and river geometries. Importantly, the ANFIS algorithm holds great potential for predicting flow fields in natural rivers because it is several orders of magnitude less expensive than a comparable simulation using LES.

This paper is organized as follows. In Section 2 we present the description of the ANFIS machine-learning algorithm. Then, in Section 3 we present the CFD model used to simulate the flow dynamics of the meandering rivers and a validation study to examine the accuracy of the CFD model for open-channel flow predictions. In Section 4, we present the test cases, and computational details of the simulated cases are presented in Section 5. Section 6 presents the results of this study. Finally, in Section 7 we conclude with the findings of the paper.

## 2 Description of ANFIS machine-learning algorithm

The Adaptive-Network-Based Fuzzy Inference System (ANFIS) is a fuzzy inference system implemented in the framework of adaptive networks [33]. It constructs a mapping from input data to output data based on "IF-THEN" rules. Take a two-input case, for example, the $i$th rule is: if $x_1$ is $A_i$ and $x_2$ is $B_i$, then $f_i = p_i x_1 + q_i x_2 + r_i$, where $A_i$ and $B_i$ are the fuzzy set, $p_i$, $q_i$ and $r_i$ are learnable parameters, $f_i$ is the result of the $i$th rule. The architecture of ANFIS is similar with a 5-layer neural network (see Figure 1, only two inputs are shown here for simplicity), but each layer has a specific function. Consider an ANFIS with $n$ inputs $x_1$, $x_2$, ..., $x_n$, the functions of each layer are introduced as follows. Layer 1 defines membership functions for each input. The membership function specifies the degree to which the input $x_i$ belongs to the fuzzy set $A_i$. This process is called fuzzification. The membership function chosen here is bell-shaped function:

$$O_{1i} = \mu_{Aj}(x_i) = \frac{1}{1+\left[\left(\frac{x_i-c_{ij}}{a_{ij}}\right)^2\right]^{b_{ij}}} \quad (1)$$

where $\mu_{Aj}$ is a membership function of fuzzy set $A$ and $a_{ij}, b_{ij}, c_{ij}$ are shape parameters, which should be manually initialized with reasonable values, and will be justified during the learning process. The index $i$ indicates the $i$th input, and $j$ indicates the $j$th membership function for each input. The number of membership functions is set up manually and can vary from each input.

Layer 2 implements the "AND" operation by multiplying the results of membership functions from different inputs. For each node in layer 2, we have:

$$O_{2i} = \Pi_i = \mu_{Aj}(x_1) \times \mu_{Bj}(x_2) \times \mu_{Cj}(x_3) \times \cdots \quad (2)$$

where $\mu_{Bj}$ and $\mu_{Cj}$ are membership functions of the fuzzy set $B$ and $C$, respectively. Layer 3 has a full connection with layer 2. It normalizes the results of layer 2 by calculating the ratio of $\Pi_i$ to the sum of all $\Pi$, as follows:

$$O_{3i} = \overline{\Pi}_i = \frac{\Pi_i}{\sum \Pi} \quad (3)$$

Layer 4 implements the "THEN" operation. Every node in this layer calculates the linear combination of all inputs and then times the normalized weight $\overline{\Pi}_i$ from layer 3:

$$O_{4i} = \overline{\Pi}_i(p_{i1}x_1 + p_{i2}x_2 + \cdots + p_{in}x_n + r_i)$$

where $p_{ij}$ and $r_i$ are the learnable parameters fine-tuned during the training process. Finally, layer 5 is the output layer. It simply sums all the outputs from layer 4, as follows:

$$O_5 = \sum O_{4i} \quad (4)$$

In this work, we employ a hybrid learning method, as follows: (1) during the training process, the parameters $a_{ij}, b_{ij}, c_{ij}$ for each membership function in layer 1 and $p_{ij}, r_i$ in layer 4 are updated by the back-propagation method and then (2) the least-squares estimate method is used to determine the parameters in layer 4 to accelerate the convergence.

## 3  The CFD model

The CFD model solves the instantaneous, incompressible, spatially-filtered Navier-Stokes equations in curvilinear coordinates. The non-dimensional form of the equations, in compact tensor notation, read as follows [42]:

$$J \frac{\partial U^j}{\partial \xi^j} = 0 \quad (5)$$

$$\frac{1}{J}\frac{\partial U^i}{\partial t} = \frac{\xi_l^i}{J}\left(-\frac{\partial}{\partial \xi^j}(U^j u_l) + \frac{1}{\rho}\frac{\partial}{\partial \xi^j}\left(\mu \frac{g^{jk}}{J}\frac{\partial u_l}{\partial \xi^k}\right) - \frac{1}{\rho}\frac{\partial}{\partial \xi^j}\left(\frac{\xi_l^i p}{J}\right) - \frac{1}{\rho}\frac{\partial \tau_{lj}}{\partial \xi^j}\right) \quad (6)$$

where $J = |\partial(\xi^1, \xi^2, \xi^3)/\partial(x_1, x_2, x_3)|$ is the Jacobian of the geometric transformation from Cartesian coordinates $\{x_i\}$ to generalized curvilinear coordinates $\{\xi^i\}$, $\xi_l^i = \partial \xi^i/\partial x_l$ are the transformation metrics, $u_i$ is the $i^{th}$ Cartesian velocity component, $U^i = (\xi_m^i/J)u_m$ is the contravariant volume flux, $g^{jk} = \xi_l^i \xi_l^k$ are the components of the contravariant metric tensor, $p$ is the pressure, $\rho$ is the fluid density, $\mu$ is the dynamic viscosity of the fluid, and $\tau_{ij}$ is the sub-grid stress tensor for LES [43]. The sub-grid stress $\tau_{ij}$ are modeled using the Smagorinsky sub-grid scale (SGS) model [44]. The governing equations are discretized in space on a hybrid staggered/non-staggered grid arrangement using second-order accurate central differencing for the convective terms and second-order accurate, three-point central differencing for the divergence, pressure gradient, and viscous-like terms [45]. The time derivatives are discretized using a second-order backward differencing scheme [43]. The discrete flow equations are integrated in time using an efficient, second-order accurate fractional step methodology coupled with a Jacobian-free, Newton-Krylov solver for the momentum equations and a GMRES solver enhanced with the multigrid method as a preconditioner for the Poisson equation.

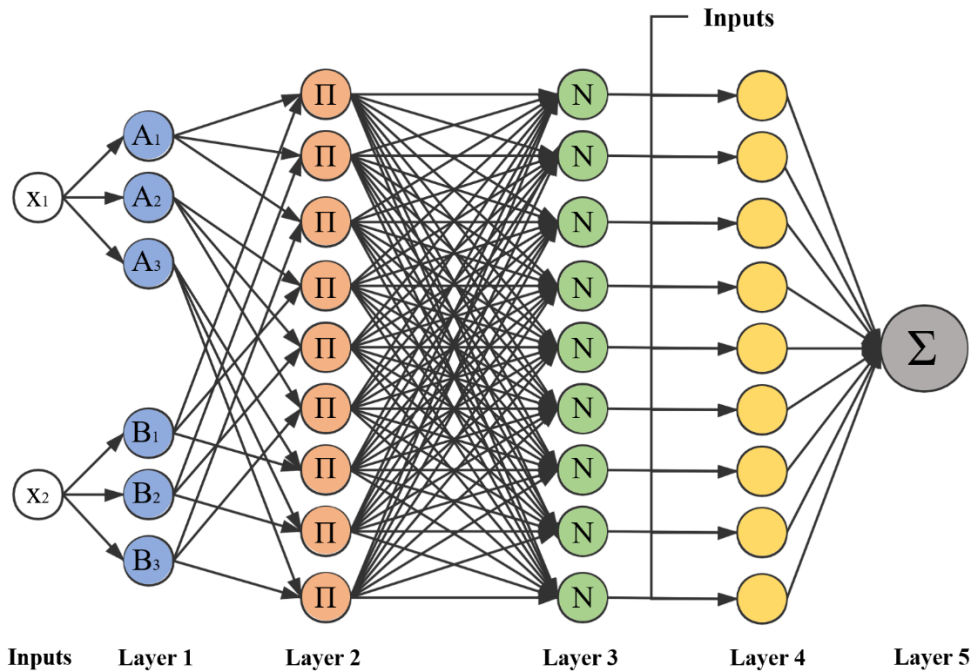

**Figure 1:** Schematic of an Adaptive-Network-Based Fuzzy Inference System (ANFIS) architecture for two inputs and one output.

The curvilinear immersed boundary method (CURVIB) is employed to handle the complex geometry of the computational domain of the large-scale meandering rivers (see Section 5) [46]. In the context of the CURVIB method, the grid nodes in the computational domain are classified into three categories: background grid nodes in the fluid phase, external nodes in the solid domains, and immersed boundary (IB) nodes, which are fluid nodes near the solid/water interfaces. The governing equations are solved in background grid nodes, while all the external nodes are blanked out from the computation. The boundary conditions are specified at IB nodes using the wall-modeling approach within the CURVIB framework [47].

## 3.1 CFD Model Validation

To validate the CFD model, we simulated a turbulent open-channel flow in a 90° bend, which was experimentally studied by Abhari et al. [48]. The experiment was carried out in a 18.6 m long, 0.6 m wide, and 0.7 m deep rectangular flume (Fig. 2). The inner and outer bend radii of the flume are 1.5 m and 2.1 m, respectively. The mean-flow depth and flow discharge are 0.2 m and 0.03 m$^3$s$^{-1}$, respectively, which results in a mean-flow velocity of 0.25 m s$^{-1}$ and Reynolds number of 5×10$^4$. A programmable electromagnetic liquid velocimeter (P-EMS) velocimeter was used to measure the streamwise and spanwise components of the velocity field within the bend.

The computational domain, which has the same geometry as the experimental flume, was split into two zones. The first zone (zone I) includes a 2.4 m long straight channel upstream that was discretized with 921×227×104 computational grid nodes in the streamwise, spanwise and vertical directions, respectively. The second zone (zone II) consists of the rest of the channel, which was discretized with 3545×227×104 computational grid nodes in the streamwise, spanwise and vertical directions, respectively. Both grids were stretched in vertical and spanwise directions so that the first node off the wall is located at a $z^+$ of about 20. A precursor LES was performed in the zone I (i.e., the straight approach channel) to obtain a fully-developed turbulent flow field [1]. In the precursor LES, the periodic boundary condition was employed in the streamwise direction. Then, the instantaneous fully-developed turbulent velocity field at the outlet cross plane of zone I was saved and used to describe the inlet boundary condition of zone II. At the outlet of zone II, we employ the Neumann boundary condition for the velocity field and turbulence quantities. The free surface in both parts was treated as a rigid-lid [49]. A non-dimensional time step (= $\Delta t \frac{U_b}{H}$, where $\Delta t$ is the physical time-step, $U_b$ is the mean-flow velocity, and $H$ is the mean-flow depth) of 0.002 was used, and LES was continued until a converged solution was obtained.

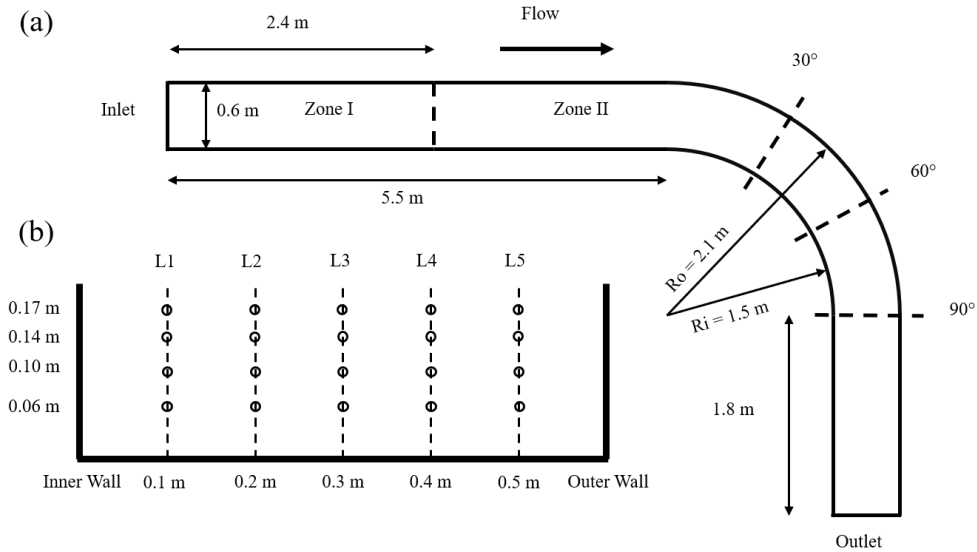

**Figure 2:** The experimental flume of Abhari et al. [48] used to validate the LES model. (a) shows the top view of the flume with a 90° bend. Measurement cross-sections of 30°, 60°, 90° are marked with dash lines. Flow field measurements are done along the dashed line at 30°, 60°, and 90° in (a). (b) shows the measurement cross-section, in which circles represent the probe locations. In (a), flow is from left to right.

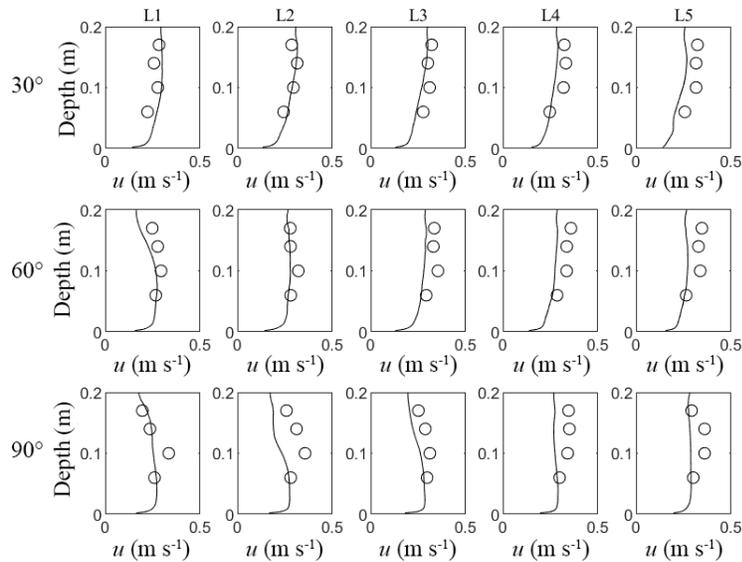

**Figure 3:** Measured (circles) and LES computed (solid lines) streamwise velocity (u) profiles at different cross-sections and along the lines L1 to L5 of Fig. 2.

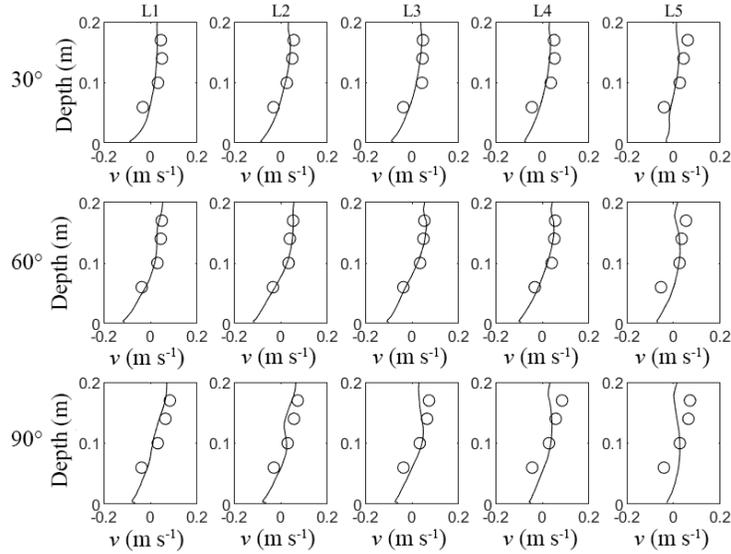

**Figure 4:** Measured (circles) and LES computed (solid lines) spanwise velocity (*v*) profiles at different cross-sections and along the lines L1 to L5 of Fig. 2.

The velocity field was measured at cross-sections of 30°, 60°, and 90°. As each cross-section, five velocity profiles, which are marked as L1, L2, L3, L4, and L5 in Fig. 2b, were measured and used for comparison with the LES results. In Figs. 3 to 4, we plot the measured and LES-computed time-averaged velocity field along the L1, L2, L3, L4, and L5 lines of Fig. 2b. As seen, the LES results for the three velocity components are in good agreement with measurements.

## 4 Virtual large-scale meandering river test-beds and the test cases

Four different river reaches (channels 1 to 4) are modeled to represent a variety of planform geometries found in meandering rivers (Fig. 5). Table 1 presents the geometrical and hydrodynamic characteristics of the virtual rivers. All four channels are constructed at scales typical of natural rivers, with channel width fixed at 100 meters, channel depth fixed at 3.3 meters, and a width-to-depth ratio of approximately 30 that is common in single-thread channels [50]. Channel bends have been distinguished between simple bends with a single dominant sense of curvature, and compound bends that that have multiple arcs of distinct curvature [51-52]. Therefore, we constructed channels with one or more bends to reflect these shapes. Sinuosity, radius of curvature, and total channel length vary across these four cases.

Channels 1 to 3 were constructed as Kinoshita curves using a common geometric model for the centerlines of meandering rivers [53-54]:

$$\theta(s) = \theta_0 \sin\left(\frac{2\pi s}{\lambda}\right) + \theta_0^3 \left(J_s \cos\left(\frac{6\pi s}{\lambda}\right) - J_f \sin\left(\frac{6\pi s}{\lambda}\right)\right) \quad (7)$$

where $\theta$ is the local direction of the channel centerline, $s$ is position along the centerline, $\lambda$ is bend wavelength, $\theta_0$ is the peak angular amplitude, $J_s$ is a skewness coefficient and $J_f$ is a flatness coefficient. For Channels 1 to 3, meander bend wavelength was fixed at 12 channel widths [55]. Channel 1 represents a single, symmetrical bend ($\theta_0 = 80°$, $J_s = 0$, and $J_f = 0$). This case is modeled with and without bridge piers. Channel 2 is a higher-amplitude, upstream-skewed, asymmetric bend ($\theta_0 = 110°$, $J_s = 0.05$, and $J_f = 0$). Channel 3 is formed by combining two bend shapes in series: a symmetric, compound bend ("I" type in Brice, 1974), followed by a straight reach with a length of 5 channel widths, and then an asymmetric bend equivalent to that in Channel 2. The compound bend is constructed using a mirrored direction sequence ($\theta_0 = 100°$, $J_s = 0$, and $J_f = 0.12$). In contrast to these Kinoshita curves, some meander bends have distinct geometries imposed by confining valley margins that limit bend amplitude (Howard, 1996). Channel 4 represents this condition using three consecutive confined meanders modeled after a bend on the Beaver River, Alberta, Canada [53, 56].

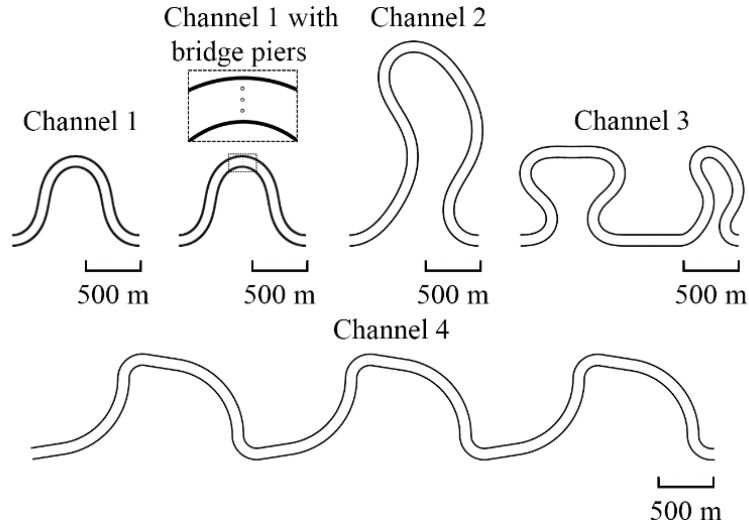

**Figure 5:** Schematics (planforms) of the virtual rivers, Channels 1 to 4, from the top view. The three virtual bridge piers in "Channel 1 with bridge piers" are cylindrical in shape with a diameter of 2 m and placed at identical distances of 25 m away from each other at the apex of Channel 1. The scale of each river is shown with a horizontal bar, while details of geometrical and hydrodynamic characteristics of the rivers are shown in Table 1.

**Table 1:** Geometrical and hydrodynamic characteristics of the virtual rivers, Channels 1 to 4. $H$, $B$, and $L$ are the mean-flow depth, width, and the total length, respectively. $S$ is the sinuosity. $U_b$ is the mean-flow velocity that is associated with the flood flow. $Re$ and $Fr$ are Reynolds and Froude numbers, both calculated based on the mean-flow depth and the bulk velocity.

|  | Channel 1 | Channel 2 | Channel 3 | Channel 4 |
| --- | --- | --- | --- | --- |
| $H$ (m) | 3.3 | 3.3 | 3.3 | 3.3 |
| $B$ (m) | 100 | 100 | 100 | 100 |
| $L$ (m) | 2110 | 4580 | 5361 | 8790 |
| $S$ | 1.76 | 3.83 | 2.71 | 1.46 |
| $U_b$ (m s$^{-1}$) | 2.04 | 2.04 | 2.04 | 2.04 |
| $Fr$ | 0.36 | 0.36 | 0.36 | 0.36 |
| $Re$ | 6.74×10$^7$ | 6.74×10$^7$ | 6.74×10$^7$ | 6.74×10$^7$ |

To examine the potential of the proposed ANFIS algorithm for generating 3D realizations of the turbulent flow in large-scale rivers, herein, we study six test cases, which are described in Table 2. It should be noted that the converged LES results of the flood flow in Channel 1 with the $Re$ number of 6.74×10$^6$ are employed to train the ANFIS algorithm, and thus we denote these LES results as "training set." Test cases 1 and 2 include Channel 1 with the $Re$ numbers of 6.74×10$^4$ and 6.74×10$^7$, respectively. Various $Re$ numbers are created by increasing the bulk velocity of the flood flow in Channel 1. Test cases 3, 4, and 5 include Channel 2, 3, and 4, respectively, with $Re$ number of 6.74×10$^6$. Test case 6 includes Channel 1 with three bridge piers, described in Figure 5 and $Re$ number of 6.74×10$^6$.

## 5 Computational details

The computational domains of the virtual rivers, Channels 1 to 4, are created to match the geometries shown in Fig. 5. The background structured grid system for each river, therefore, contains the fluid (water) phase of that river. While the side-walls and the flatbed of each channel are created using unstructured triangular grid systems and immersed into the water phased using the CURVIB approach. Details of each grid system used to model the flood flow of the virtual rivers are presented in Table 3.

**Table 2:** Description of test cases 1 to 6. LES results of Channel 1 with $Re$ number of $6.74\times10^6$ constitute the training set of our ANFIS machine-learning algorithm.

| Test case | Test-bed | Re number |
|---|---|---|
| Training set | Channel 1 | $6.74\times10^6$ |
| Test Case 1 | Channel 1 | $6.74\times10^4$ |
| Test Case 2 | Channel 1 | $6.74\times10^7$ |
| Test Case 3 | Channel 2 | $6.74\times10^6$ |
| Test Case 4 | Channel 3 | $6.74\times10^6$ |
| Test Case 5 | Channel 4 | $6.74\times10^6$ |
| Test Case 6 | Channel 1 with piers | $6.74\times10^6$ |

LESs of the flood flows in the large-scale virtual rivers are carried out using periodic boundary conditions in the streamwise direction, while the free surface of the rivers is described using the rigid-lid assumptions. For each river, the flow rate of the flood flow is prescribed at the inlet cross-section (see Table 1). The LES of each river is continued until the flow field statistically converges. We identified convergence by monitoring the evolution of the total kinetic energy of the flow of each river. The so-obtained fully converged LES results are used to train and validate the ANFIS machine-learning algorithm.

**Table 3:** Computational details of the background grid system of the four virtual rivers, Channels 1 to 4. $N_x$, $N_y$, and $N_z$ are the number of computational grid nodes in streamwise, spanwise, and vertical directions, respectively. $\Delta x$, $\Delta y$, and $\Delta z$ are the special resolution in streamwise, spanwise, and vertical directions, respectively. $z^+$ is the vertical resolution in the wall unit and $\Delta t$ is the temporal resolution.

| | Channel 1 | Channel 2 | Channel 3 | Channel 4 |
|---|---|---|---|---|
| $N_x \times N_y \times N_z$ | $2201 \times 121 \times 21$ | $4613 \times 121 \times 21$ | $6251 \times 121 \times 21$ | $10713 \times 121 \times 21$ |
| $\Delta x$ (m) | 0.96 | 0.99 | 0.86 | 0.82 |
| $\Delta y$ (m) | 0.83 | 0.83 | 0.83 | 0.83 |
| $\Delta z$ (m) | 0.17 | 0.17 | 0.17 | 0.17 |
| $z^+$ | 13000 | 13000 | 13000 | 13000 |
| $\Delta t$ (s) | 0.08 | 0.08 | 0.08 | 0.08 |

# 6 Results and discussion

ANFIS has been shown to have a great potential in predicting time-series of variables for fluid dynamics applications [57–61]. To explore this potential, we consider an evenly-distributed discrete time series of variables $f(1)$ $f(2)$, …, $f(t)$. ANFIS is very well suited to create a mapping from known previous variables (e.g., $f(t-3), f(t-2), f(t-1)$, and $f(t)$) onto an unknown future variable (e.g., $f(t+1)$). Instantaneous turbulent flow fields obtained from LES with a constant time-step size, $\Delta t$, are even-distributed discrete time-series, as well. Using this characteristic, in this work, we will use ANFIS to forecast 3D turbulent flow velocity components (Cartesian) of $u$, $v$, and $w$ at the time-step $t+1$ using the LES results at time-steps $t, t-1, t-2$, and $t-3$. Ideally, if the predicted variables are recycled as inputs in the future times, ANFIS should be able to keep forecasting the next time-steps forever. Data-driven machine-learning algorithms, however, are essentially based on curve-fitting methods, and the error in each step of prediction accumulates and/or magnifies. In practice, such accumulation of error eventually leads to the restriction of the predictive ability of machine-learning algorithms.

In this paper, we train an ANFIS using the training set (Table 2) to predict flood flow velocity field of the six test cases (Table 2), i.e., the large-scale virtual rivers. The instantaneous LES results of the flood flow in the test-bed of Channel 1 with the $Re$ number of 6.74×10$^6$ constitute our training set to fine-tune the learnable coefficients of the ANFIS machine-learning algorithm. We also ran separately LESs to generate CFD-based instantaneous flood flow velocity fields for the six test cases to validate the ANFIS predictions and to examine the accuracy of its predictions. We utilized three different statistical error indices, such as coefficient of determination ($R^2$), mean absolute error (MAE), root mean square error (RMSE), and mean absolute relative error (MARE) to evaluate the accuracy of the ANFIS predictions. These statistical error indices are defined as follows [62]:

$$R^2 = 1 - \frac{\sum_{i=1}^{N}(\psi_{i(ANFIS)}-\psi_{i(LES)})^2}{\sum_{i=1}^{N}(\psi_{i(ANFIS)}-\overline{\psi}_{i(ANFIS)})^2} \tag{8}$$

$$MAE = \frac{\sum_{i=1}^{N}|\psi_{i(ANFIS)}-\psi_{i(LES)}|}{N} \tag{9}$$

$$RMSE = \left(\frac{\sum_{i=1}^{N}(\psi_{i(ANFIS)}-\psi)^2}{N}\right)^{0.5} \tag{10}$$

$$MARE = \frac{1}{N}\sum_{i=1}^{N}\frac{|\psi_{i(ANFIS)}-\psi_{i(LES)}|}{\psi_{i(LES)}} \tag{11}$$

where $\psi_{i(ANFIS)}$ is the predicted value using the ANFIS machine-learning algorithm, $\psi_{i(LES)}$ is the value obtained using the LES model, $\bar{\psi}_{i(ANFIS)}$ is the mean predicted value using the ANFIS, and $N$ is the total number of samples, i.e., the total number of computational nodes to discretize the flow domains of the large-scale rivers.

As expected, the computational cost of LES for the selected test cases (i.e., the large-scale rivers test-beds of Channel 1 to 4) is several times higher than that of the ANFIS predictions. For example, the LES of a single time-step of the flow field in Channel 1 of the test case 1 takes about 180 s to complete on a single CPU. In contrast, the required time for the same computations with an ANFIS is approximately 24 s. This comparison illustrates how advantageous such machine-learning algorithms can be for predicting 3D realizations of the flow field in large-scale natural rivers.

## 6.1 Training the ANFIS machine-learning algorithm

The training set data were obtained from fully converged LES results of Channel 1 with $Re$ number of $6.74 \times 10^6$ (Table 2). The computational grid system of Channel 1 includes roughly 5.6 million grid nodes (Table 3). The training set data consist of the three velocity components $u$, $v$, and $w$ on 10,000 computational nodes, out of roughly 5.6 million. The computational nodes for the trained set data were selected randomly and uniformly. The training set data of the velocity field were selected from five successive time steps, which are denoted as $t_1$, $t_2$, $t_3$, $t_4$, and $t_5$. During the training process of the ANFIS algorithm, the instantaneous velocity field at times $t_1$, $t_2$, $t_3$, and $t_4$ were treated as the input arrays while the instantaneous velocity field at $t_5$ served as the target or the output array.

The training process can be done using either of the three velocity components at times $t_1$ to $t_5$. We trained three different ANFIS algorithms, each using one of the velocity components. The trained ANFIS algorithm using the streamwise, spanwise, and vertical velocity components are denoted as ANFIS$_u$, ANFIS$_v$, and ANFIS$_w$, respectively. Then, we check the predictive capabilities of each of those trained algorithms by comparing their predicted velocity components (i.e., all three velocity components of $u$, $v$, and $w$) at time $t_5$ with those of the LES results. We note that two of the velocity components were not introduced to the ANFIS algorithms during their training process. The comparison between the three trained algorithms and the LES results are shown in Figure 6. This figure shows that the ANFIS$_v$ algorithm displays the best performance among the three algorithms for predicting not only $v$ but also the other two velocity components of $u$ and $w$ with excellent accuracy. The ranges of LES computed non-dimensional streamwise, spanwise, and vertical velocity components ($u/U_b$, $v/U_b$, $w/U_b$, respectively) in Channel 1 are -0.2 < $u$ < 1.8, -1.8 < $v$ < 1.8, and -0.3 < $w$ < 0.3, respectively. Given the variation of the velocity components, the good performance of the ANFIS$_v$ algorithm could be attributed to the wide range of spanwise velocity component, which spans the range of the streamwise and vertical velocity components. Therefore, the range of the velocity component selected to train the machine-learning algorithm clearly plays an important role in the performance of the trained algorithm. As a result, all the test cases were performed in the V model. Given the performance of the ANFIS$_v$ algorithm, in the rest of this paper, we will use it to produce 3D realizations of the flood flow field in all other test cases and denote it as the "trained ANFIS".

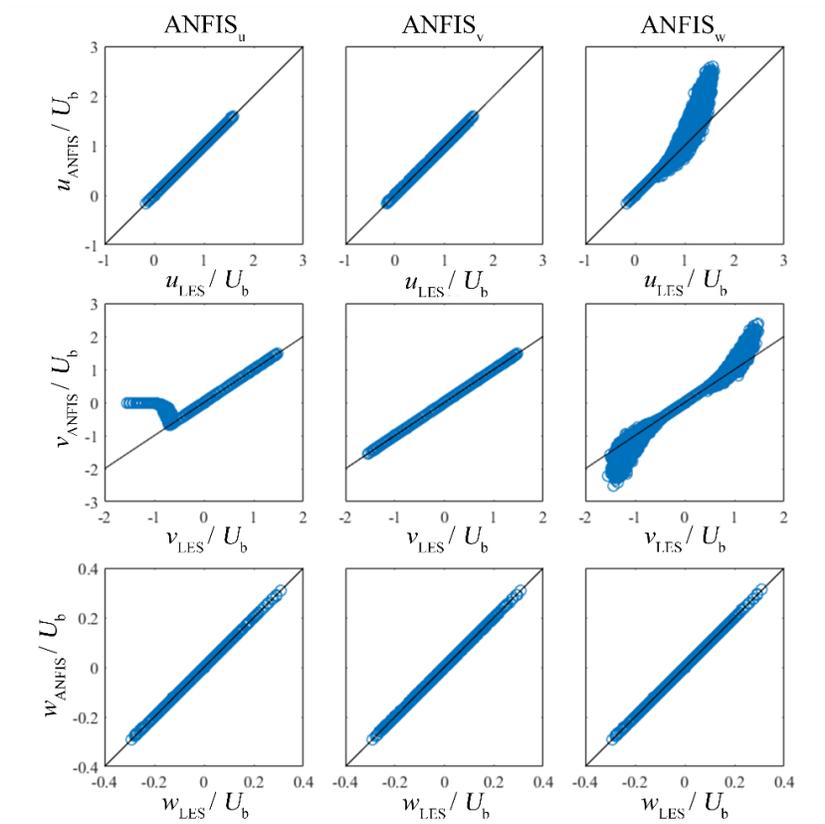

**Figure 6:** Comparing the predictive capabilities of the trained ANFIS$_u$, ANFIS$_v$, and ANFIS$_w$ algorithms with that of the LES model for prediction of the three velocity components of the flood flow in Channel 1 at time $t_5$ for the randomly selected 10,000 computational nodes. Black lines indicate the location of the perfect match between the LES results and the ANFIS algorithm. $u_{LES}$, $v_{LES}$, and $w_{LES}$, are LES computed velocity components, and $u_{ANFIS}$, $v_{ANFIS}$, and $w_{ANFIS}$ are obtained via ANFIS.

## 6.2 Prediction of one time-step march in time using the trained ANFIS algorithm

To generate 3D realizations of the flood flow field of the test cases 1 to 6 (Table 2), we first carried out LES of each test case to produce their fully converged velocity fields. Once fully converged, the LES of each test case is continued for five time-steps to produce their 3D velocity field at times $t_1$, $t_2$, $t_3$, $t_4$, and $t_5$. The 3D flow field data in the entire flow domain at the first four times of $t_1$ to $t_4$ are then stored as precursor input data for the trained ANFIS to generate the 3D flow field data at the time $t_5$. In the meantime, the LES flow field data at the time $t_5$ will be later used to validate the ANFIS generate flow field at the time $t_5$.

In Figure 7 to Figure *12*, we plot the contours of instantaneous (at time $t_5$) velocity magnitude at the free surface of large-scale rivers in test cases 1 to 6, respectively. The instantaneous results in these figures are obtained from the trained ANFIS and the LES model. Additionally, in each of these figures, we present the velocity magnitude profiles in spanwise directions in order to make a quantitative comparison between the ANFIS and LES model predictions. As seen, the ANFIS predictions for one time-step closely resemble the LES results -- even though the test cases represent meandering river flows that are starkly different, in terms of Reynolds number and/or river geometry, from the those in Channel 1 on which the ANFIS was trained. Test case 6, in particular, could be of interest to study the impact of flood flow on the stability of bridge piers' foundations. The ANFIS seems to enable computationally affordable predictions of complex turbulent flood flow around the bridge piers in this test case (Figure 12). In other words, given the high cost of conducting high-fidelity LES of the flood flow in large-scale rivers with wall-mounted hydraulic structures [1–6], such machine-learning algorithms can be a breakthrough for affordable predictions of scour around bridge pier foundations in natural rivers.

Additional quantitative comparisons are made by computing the $R^2$, MAE, RMSE, and MARE of the ANFIS and LES model predictions for the 3D flow field of the six test cases (see Table 4). As seen in this table, the ANFIS generates flow fields that fit the target LES results with excellent accuracy. This indicates that, instead of fitting a solution at a specific set of parameters, the trained ANFIS has successfully learned the non-linear dynamics of the governing equations. Furthermore, we plot in Figure 13 and Figure *14* the statistics of the entire computational grid nodes for the two representative test cases of 2 and 5 in term of differential error between the ANFIS predictions and LES results, i.e., $err = \psi_{ANFIS} - \psi_{LES}$, where $\psi_i$ represents the three velocity components of $u$, $v$, and $w$ obtained from LES model and the trained ANFIS algorithm.

As seen in Figure 13 (a-c) and Figure *14* (a-c), the distribution of the number of computational nodes in error-velocity space shows that most of the computational nodes correspond to an error of $err \approx 10^{-8}$. In other words, regardless of the range of the velocity components, the ANFIS predictions for most of the computational nodes contain negligible level of errors. As seen in these figures, computational nodes with higher velocity (in absolute value) at the two extreme ends of the distributions correspond to a broader range of errors (i.e., the expansion of the data in vertical). At the same time, the distributions of computational nodes with high velocities at the two extreme ends of the data seem more scattered in error because the total number of computational nodes at the high-velocity range is limited. This observation appears to be valid for all three velocity components.

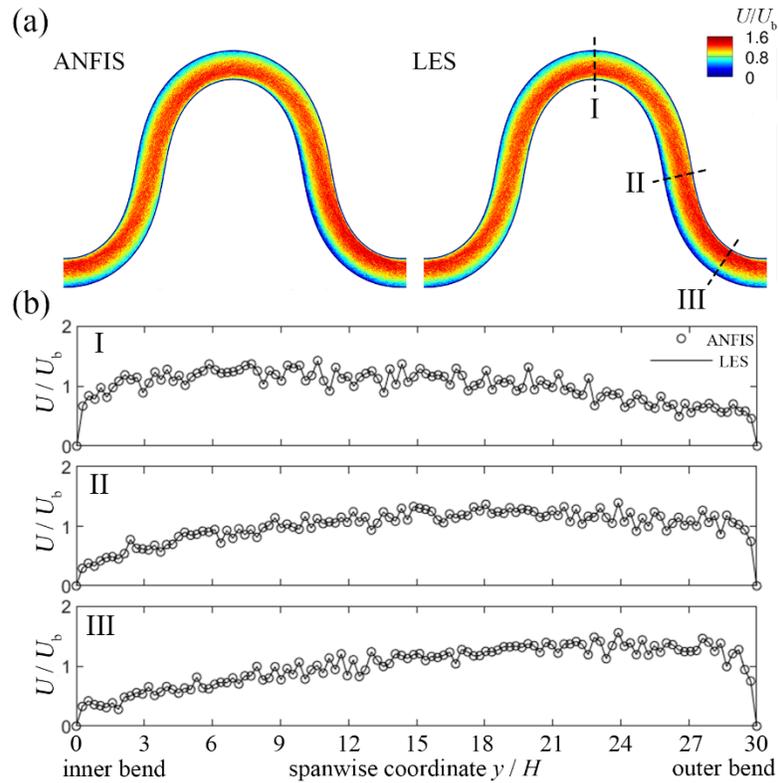

**Figure 7:** Instantaneous LES results and ANFIS predictions for the 3D flow field of the test case 1 at time $t_5$. (a) shows the contours of velocity magnitude ($U / U_b$) at the free surface of Channel 1 from the top view. In (a), flow is from left to right. (b) shows the profiles of the velocity magnitude in the spanwise direction along the three dashed lines of I, II, and III in (a). In (b), solid lines represent the LES results, while circles represent the predictions of ANFIS.

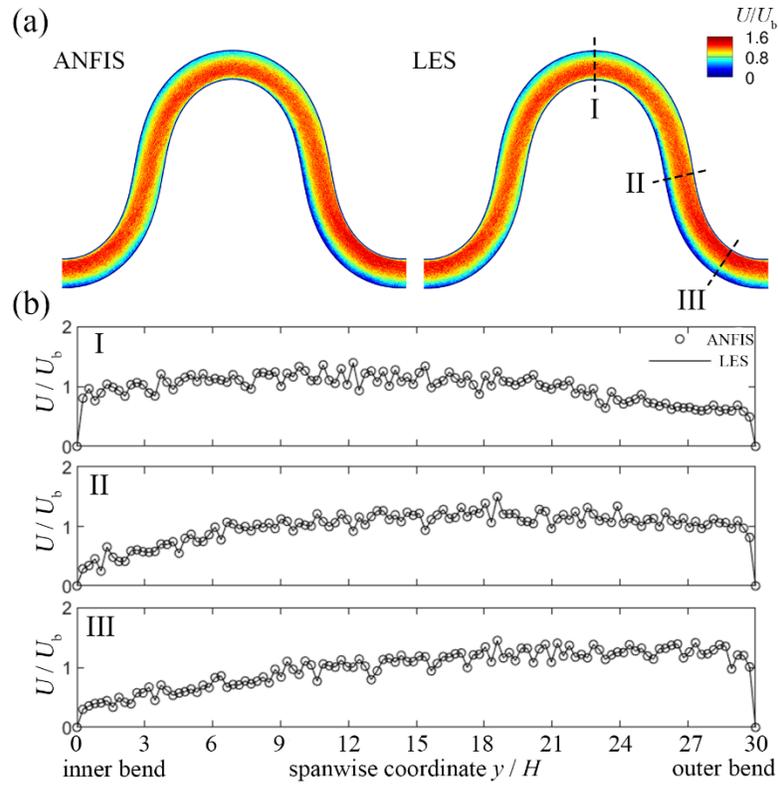

**Figure 8:** Instantaneous LES results and ANFIS predictions for the 3D flow field of the test case 2 at time $t_5$. (a) shows the contours of velocity magnitude ($U / U_b$) at the free surface of Channel 1 from the top view. In (a), flow is from left to right. (b) shows the profiles of the velocity magnitude in the spanwise direction along the three dashed lines of I, II, and III in (a). In (b), solid lines represent the LES results, while circles represent the predictions of ANFIS.

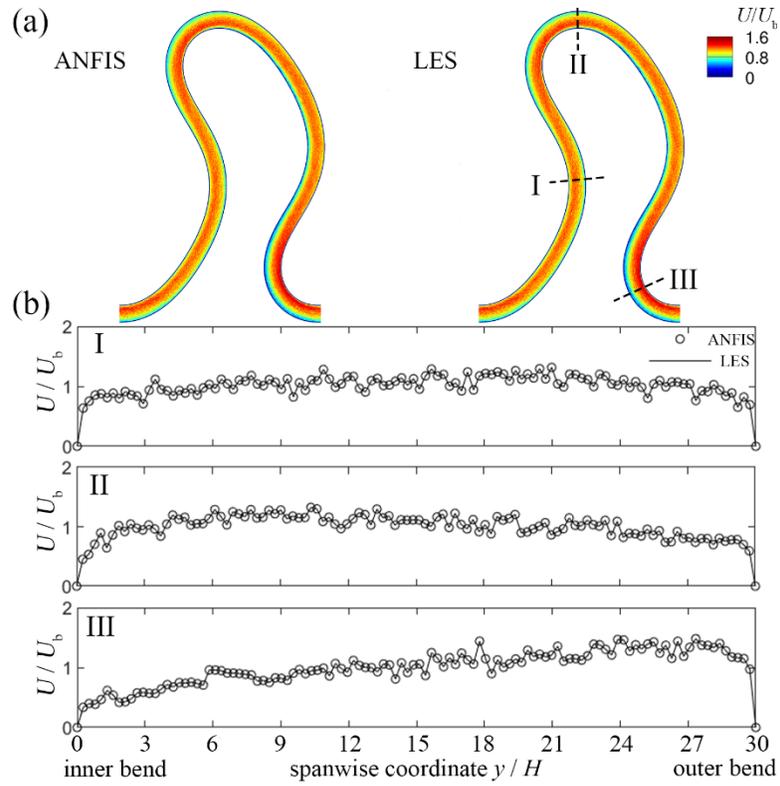

**Figure 9:** Instantaneous LES results and ANFIS predictions for the 3D flow field of the test case 3 at time $t_5$. (a) shows the contours of velocity magnitude ($U / U_b$) at the free surface of Channel 2 from the top view. In (a), flow is from left to right. (b) shows the profiles of the velocity magnitude in the spanwise direction along the three dashed lines of I, II, and III in (a). In (b), solid lines represent the LES results, while circles represent the predictions of ANFIS.

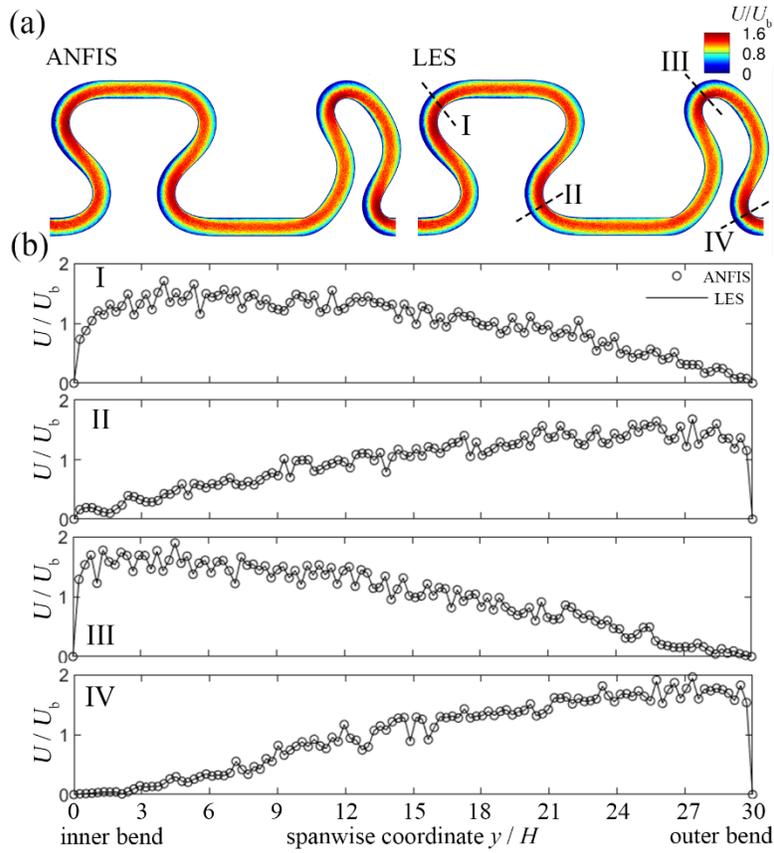

**Figure 10:** Instantaneous LES results and ANFIS predictions for the 3D flow field of the test case 4 at time $t_5$. (a) shows the contours of velocity magnitude ($U / U_b$) at the free surface of Channel 3 from the top view. In (a), flow is from left to right. (b) shows the profiles of the velocity magnitude in the spanwise direction along the three dashed lines of I, II, and III in (a). In (b), solid lines represent the LES results, while circles represent the predictions of ANFIS.

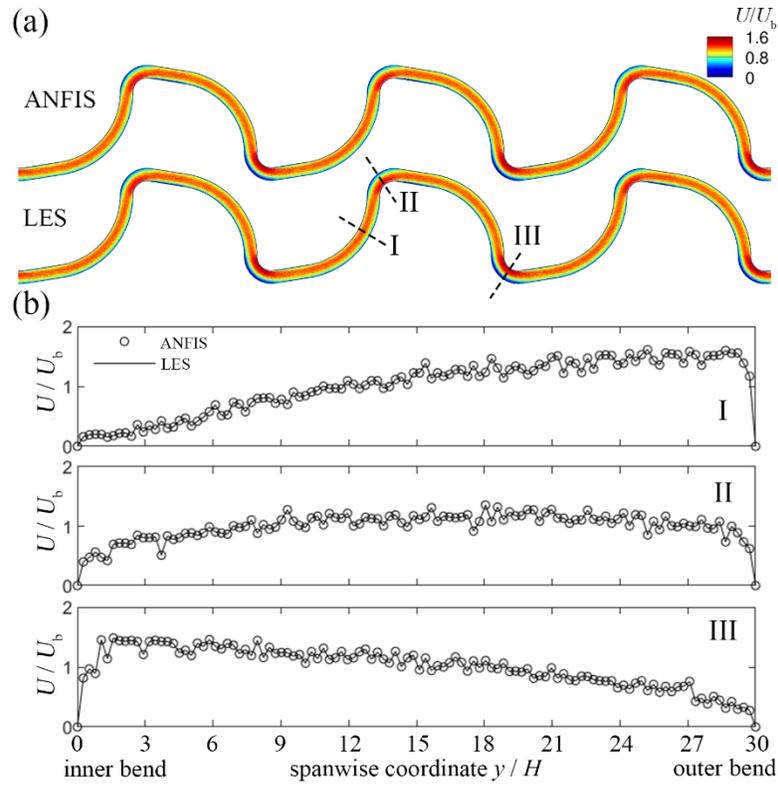

**Figure 11:** Instantaneous LES results and ANFIS predictions for the 3D flow field of the test case 5 at time $t_5$. (a) shows the contours of velocity magnitude ($U / U_b$) at the free surface of Channel 4 from the top view. In (a), flow is from left to right. (b) shows the profiles of the velocity magnitude in the spanwise direction along the three dashed lines of I, II, and III in (a). In (b), solid lines represent the LES results, while circles represent the predictions of ANFIS.

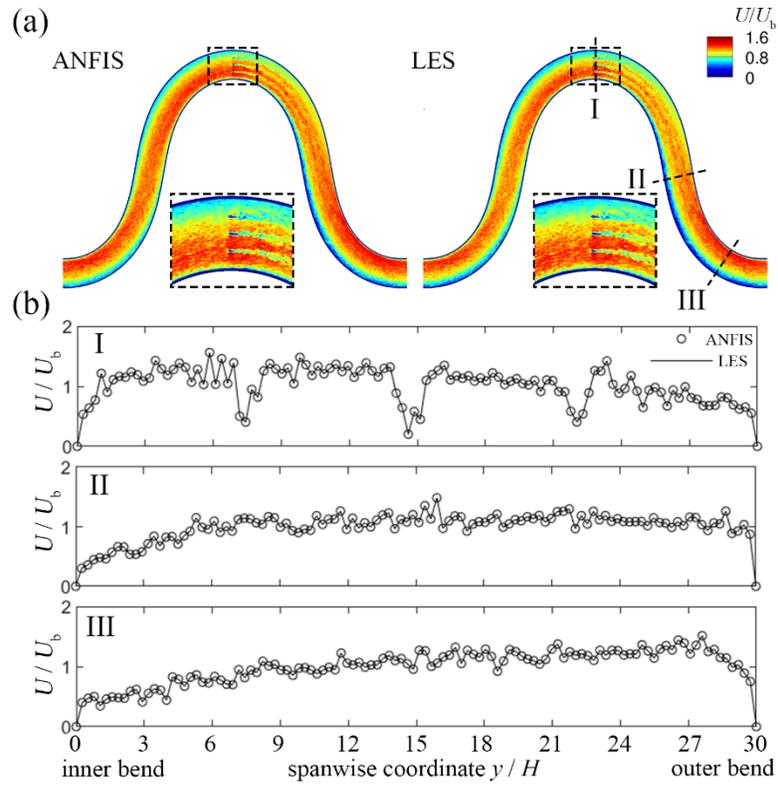

**Figure 12:** Instantaneous LES results and ANFIS predictions for the 3D flow field of the test case 6 at time $t_5$. (a) shows the contours of velocity magnitude ($U / U_b$) at the free surface of Channel 1 (with three bridge piers) from the top view. In (a), flow is from left to right. (b) shows the profiles of the velocity magnitude in the spanwise direction along the three dashed lines of I, II, and III in (a). In (b), solid lines represent the LES results, while circles represent the predictions of ANFIS.

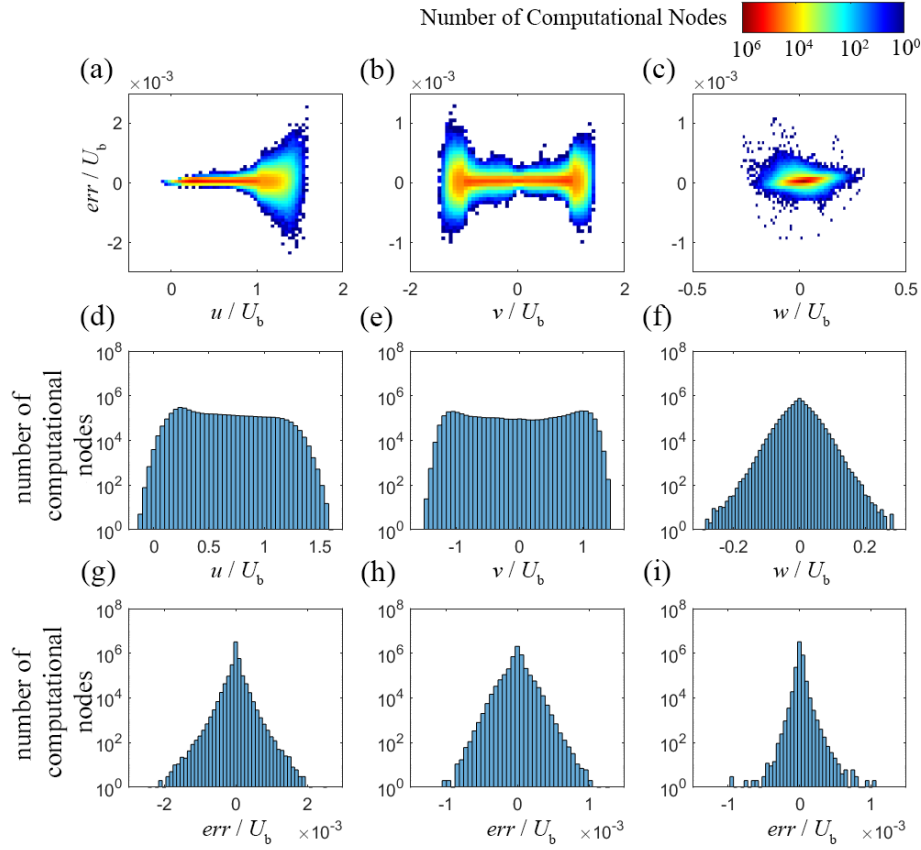

**Figure 13:** Error distribution of the ANFIS predictions relative to the LES results for test case 2. (a) to (c) illustrate the number of computational nodes in the error-velocity space for the streamwise ($u$), spanwise ($v$), and vertical ($w$) velocity components, respectively. (d) to (f) depict the distribution of the number of computational nodes as a function of their corresponding velocity range. (g) to (i) plot the distribution of computational nodes as a function of their error range.

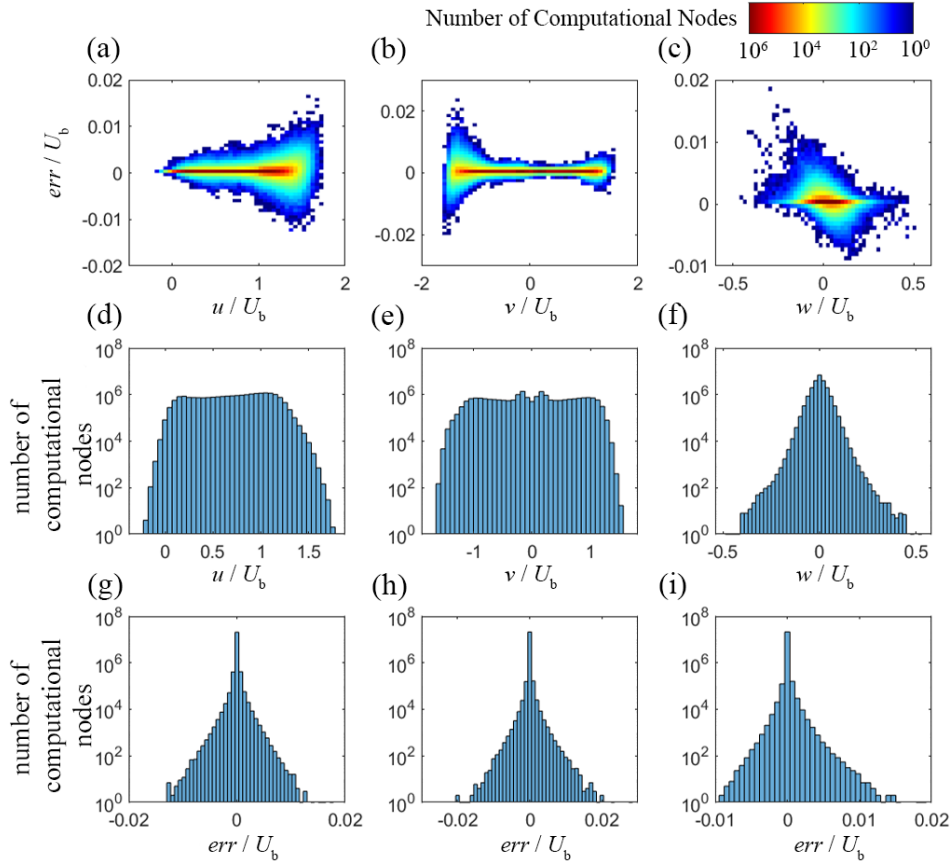

**Figure 14:** Error distribution of the ANFIS predictions relative to the LES results for test case 5. (a) to (c) illustrate the number of computational nodes in the error-velocity space for the streamwise ($u$), spanwise ($v$), and vertical ($w$) velocity components, respectively. (d) to (f) depict the distribution of the number of computational nodes as a function of their corresponding velocity range. (g) to (i) plot the distribution of computational nodes as a function of their error range.

**Table 4:** Statistical error indices for one time-step march in time predictions of the ANFIS relative to the LES results for test cases 1 to 6. $R^2$ is the coefficient of determination, MAE is the mean absolute error, RMSE is the root mean square error, and MARE is the mean absolute relative error (see Eqns. 7 to 10).

|             | $R^2$ | MAE | RMSE | MARE |
|---|---|---|---|---|
| Test case 1 | 1.00 | $7.93 \times 10^{-5}$ | $1.38 \times 10^{-4}$ | $7.18 \times 10^{-5}$ |
| Test case 2 | 1.00 | $8.86 \times 10^{-5}$ | $1.54 \times 10^{-4}$ | $8.12 \times 10^{-5}$ |
| Test case 3 | 1.00 | $9.53 \times 10^{-5}$ | $1.87 \times 10^{-4}$ | $8.46 \times 10^{-5}$ |
| Test case 4 | 1.00 | $7.57 \times 10^{-4}$ | $2.17 \times 10^{-3}$ | $4.97 \times 10^{-5}$ |
| Test case 5 | 1.00 | $1.74 \times 10^{-4}$ | $5.41 \times 10^{-4}$ | $1.32 \times 10^{-5}$ |
| Test case 6 | 1.00 | $9.97 \times 10^{-5}$ | $2.05 \times 10^{-4}$ | $8.86 \times 10^{-5}$ |

## 6.3 Prediction of multi time-step march in time using the trained ANFIS algorithm

In this section, we attempt to extend the prediction period of the trained ANFIS to march in time beyond a single time-step. To do so, we recycle the output(s) (predictions) of the ANFIS and use them as the inputs to the ANFIS so it can predict new time steps recurrently. As described in the previous section, for a single time-step march in time, we use the LES results of the initial four time-steps, say $t_1^{LES}$, $t_2^{LES}$, $t_3^{LES}$, and $t_4^{LES}$, as inputs to the trained ANFIS to predict the 3D flow field at the fifth time-step, i.e., $t_5^{ANFIS}$. To extend the ANFIS predictions to the next time-step forward, we recycle $t_5^{ANFIS}$ and use it; along with the three prior time steps of LES results, i.e., $t_2^{LES}$, $t_3^{LES}$, and $t_4^{LES}$; as inputs to the ANFIS to predict the flow field at the time $t_6^{ANFIS}$. As this process continues, soon all input. In other words, the flow field predictions of ANFIS at times $t_{n-3}^{ANFIS}$, $t_{n-2}^{ANFIS}$, $t_{n-1}^{ANFIS}$, and $t_n^{ANFIS}$ are used as inputs to march in time and generate the 3D flow field at the time $t_n^{ANFIS}$. This procedure is shown in Figure 15. This process can ideally continue recurrently for long periods to generate 3D flow field realizations. However, as the march in time continues, the ANFIS prediction error accumulates to a point where the ANFIS predicted flow field deviates from the LES results.

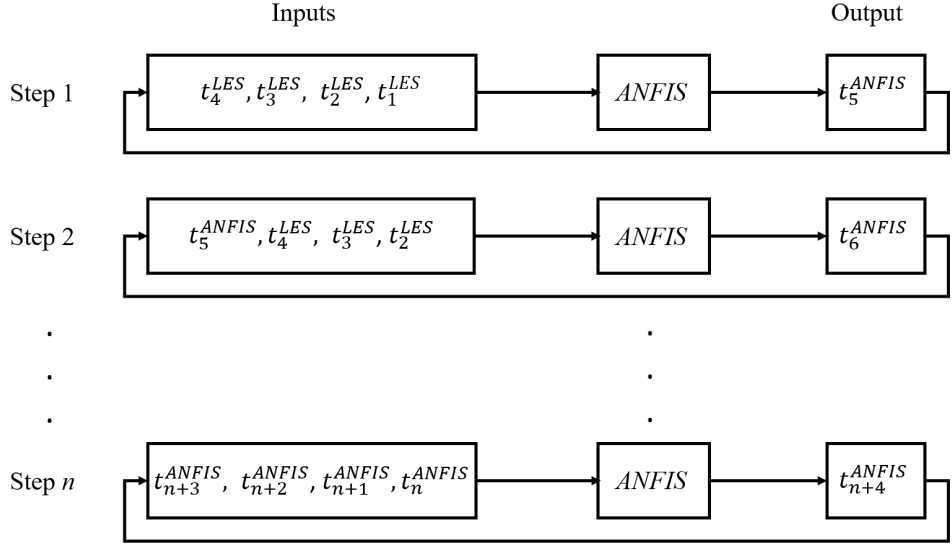

**Figure 15:** Schematics of the multiple time-step march-in-time predictions using ANFIS. $\psi_i^{LES}$ and $\psi_i^{ANFIS}$ are the LES-computed and ANFIS predicted flow field at the time $t_i$, respectively.

The flow field predictions of ANFIS for test case 1 at times $t_{10}$, $t_{20}$, $t_{30}$, and $t_{40}$ are shown in Figure 16 to Figure *19*, respectively, while their error indices are presented in Table 5. As seen, the 3D flow field realizations of ANFIS at times $t_{10}$, $t_{20}$, $t_{30}$, and $t_{40}$ are in reasonable agreement with those of the LES models. However, it can be clearly seen that the discrepancies between the ANFIS algorithm predictions and the LES-computed flow velocity increase rapidly at times greater than $t_{20}$. More specifically, comparing with the LES-computed velocity field, the ANFIS seem to overestimate the velocity field at instants of $t_{30}$, and $t_{40}$.

Given the outstanding performance of ANFIS predicting the flow field for 20 time-steps, it seems reasonable to use the trained ANFIS and LES model recurrently to increase the accuracy of results and, at the same time, reduce the cost of flow field predictions. We note that in this recurrent approach, like the approach discussed above, it is not required to re-train the ANFIS for each cycle of predictions. More specifically, the trained ANFIS can be used for predicting the flow field until $t_{20}$. Then, using the instantaneous results of ANFIS at $t_{20}$, LES model can be run to simulate four time-steps, i.e., until time $t_{24}$. Subsequently, the four time-steps of data from LES are used as the precursor (inputs) time-steps by the ANFIS to predict 20 time-steps of the flow field, i.e., until $t_{44}$. Ideally, this recurrent procedure can continue for long enough periods of time to generate enough data for a statistically converged time-averaged solution. We note that, since only four time-steps of LES are needed after each 20 time-steps, the computational cost of LES will not be significant. To demonstrate the promise of this recurrent approach, we conducted LES of ANFIS-predicted flow field (for test case 1) at time $t_{20}$ for four time-steps. The four time-steps of LES generated flow field then were used as the precursor inputs to the ANFIS to predict another 20 time-steps of 3D flow field until $t_{44}$. Then, four more time-steps of LES and 20 more time-steps of ANFIS were conducted to predict the 3D flow field until $t_{68}$. In order to distinguish this approach from the direct ANFIS approach, herein, we denote it as the LES-ANFIS hybrid method. Figure 20 compares the so-predicted LES-ANFIS results at the time $t_{68}$ with those of LES. As seen in this figure, the proposed hybrid LES-ANFIS approach has the potential to successfully generate 3D realization of the velocity field of flood flow in the large-scale river for long periods of time.

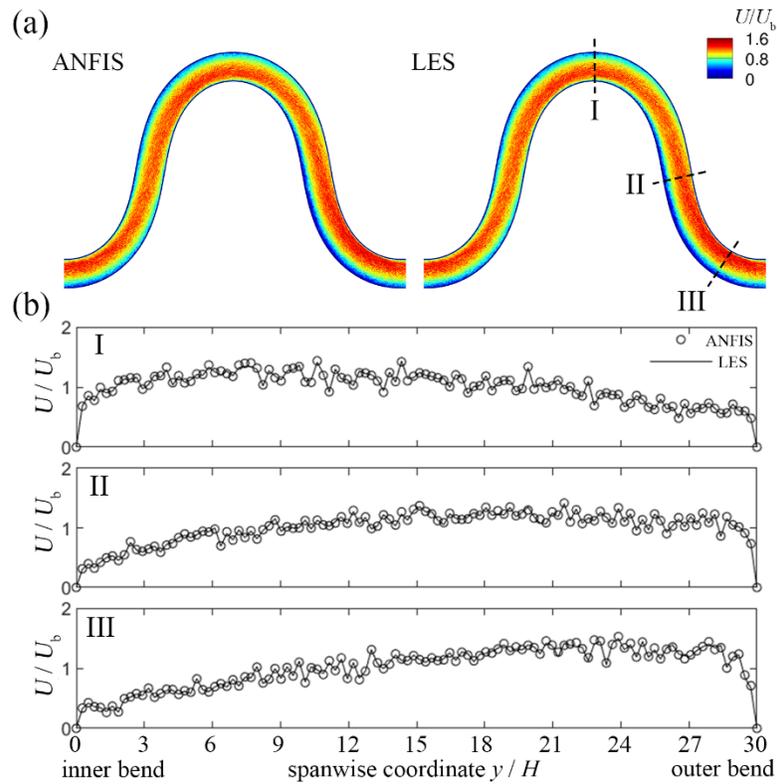

**Figure 16:** Instantaneous LES results and ANFIS predictions for the 3D flow field of the test case 1 at time $t_{10}$. (a) shows the contours of velocity magnitude ($U / U_b$) at the free surface of Channel 1 from the top view. In (a), flow is from left to right. (b) shows the profiles of the velocity magnitude in the spanwise direction along the three dashed lines of I, II, and III in (a). In (b), solid lines represent the LES results, while circles represent the predictions of ANFIS.

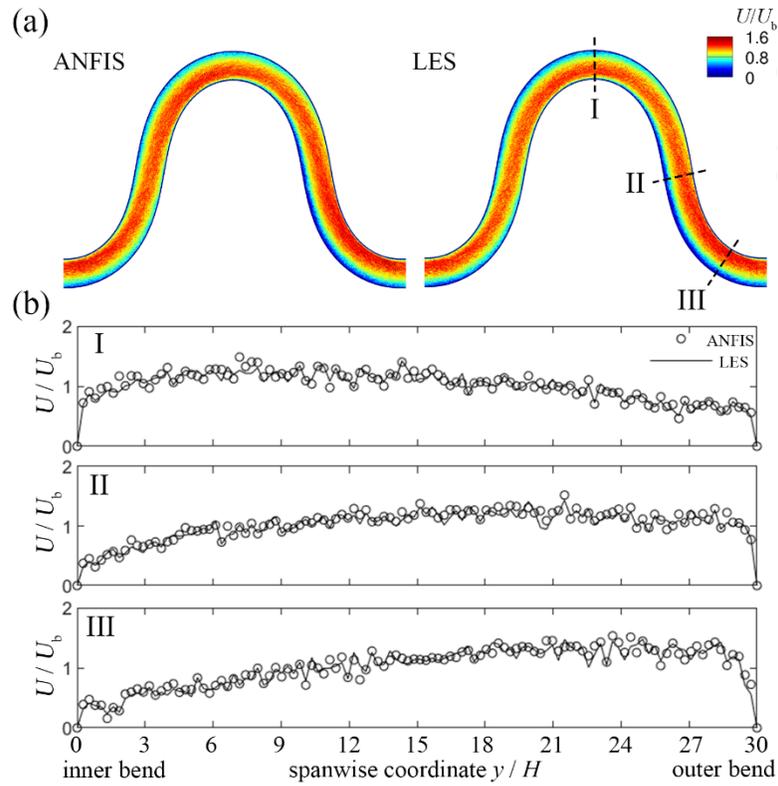

**Figure 17:** Instantaneous LES results and ANFIS predictions for the 3D flow field of the test case 1 at time $t_{20}$. (a) shows the contours of velocity magnitude ($U/U_b$) at the free surface of Channel 1 from the top view. In (a), flow is from left to right. (b) shows the profiles of the velocity magnitude in the spanwise direction along the three dashed lines of I, II, and III in (a). In (b), solid lines represent the LES results, while circles represent the predictions of ANFIS.

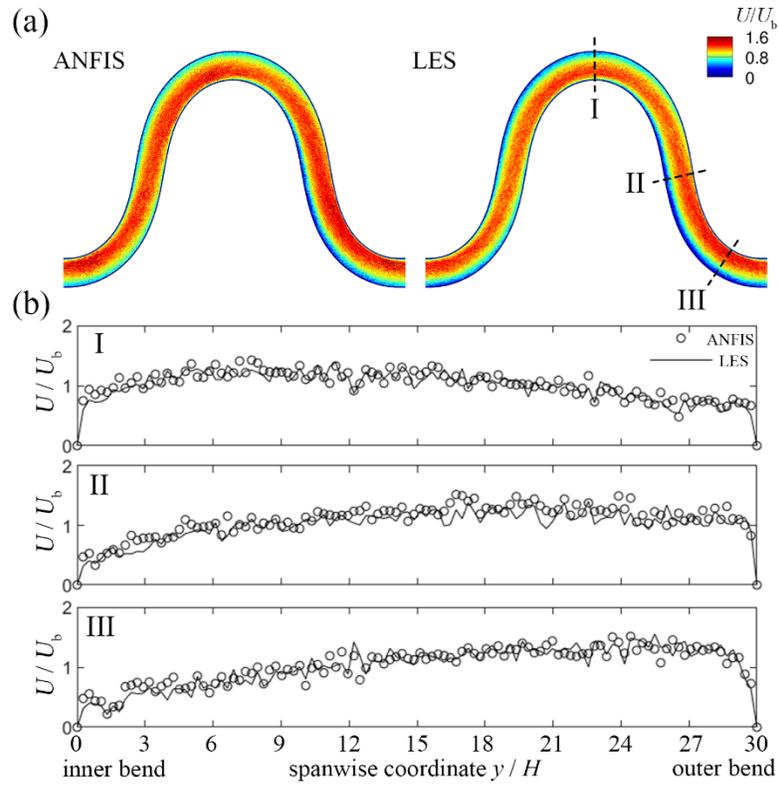

**Figure 18:** Instantaneous LES results and ANFIS predictions for the 3D flow field of the test case 1 at time $t_{30}$. (a) shows the contours of velocity magnitude ($U/U_b$) at the free surface of Channel 1 from the top view. In (a), flow is from left to right. (b) shows the profiles of the velocity magnitude in the spanwise direction along the three dashed lines of I, II, and III in (a). In (b), solid lines represent the LES results, while circles represent the predictions of ANFIS.

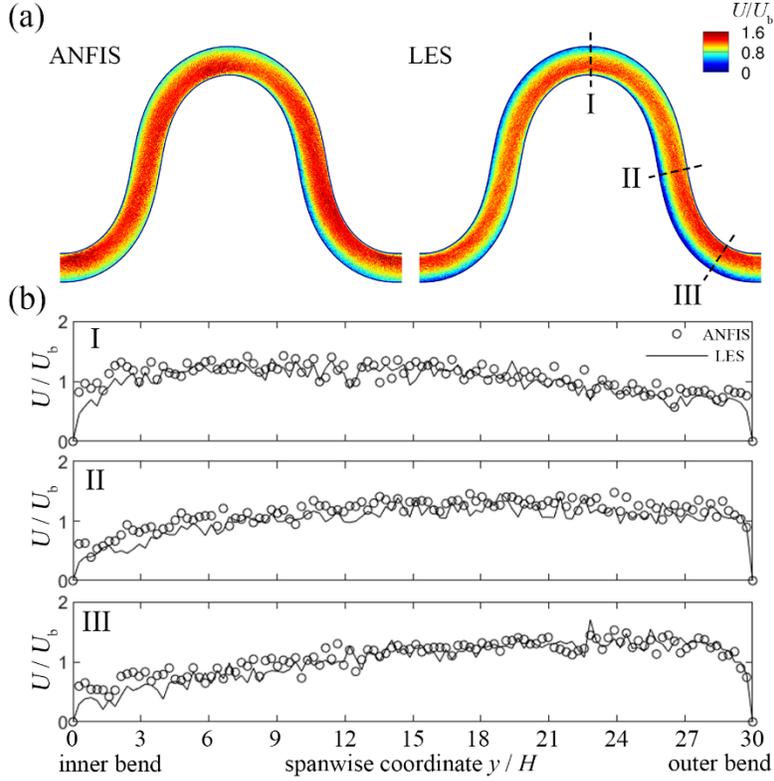

**Figure 19:** Instantaneous LES results and ANFIS predictions for the 3D flow field of the test case 1 at time $t_{40}$. (a) shows the contours of velocity magnitude ($U / U_b$) at the free surface of Channel 1 from the top view. In (a), flow is from left to right. (b) shows the profiles of the velocity magnitude in the spanwise direction along the three dashed lines of I, II, and III in (a). In (b), solid lines represent the LES results, while circles represent the predictions of ANFIS.

**Table 5:** Statistical error indices for multiple time-step march-in-time predictions of the ANFIS relative to the LES results for test cases 1. $R^2$ is the coefficient of determination, MAE is the mean absolute error, RMSE is the root mean square error, and MARE is the mean absolute relative error (see Eqns. 7 to 10).

| Time | $R^2$ | MAE | RMSE | MARE |
|---|---|---|---|---|
| $t_{10}$ | 1.00 | 0.01 | 0.02 | 0.01 |
| $t_{20}$ | 0.93 | 0.06 | 0.09 | 0.06 |
| $t_{30}$ | 0.79 | 0.10 | 0.13 | 0.11 |
| $t_{40}$ | 0.63 | 0.13 | 0.14 | 0.16 |

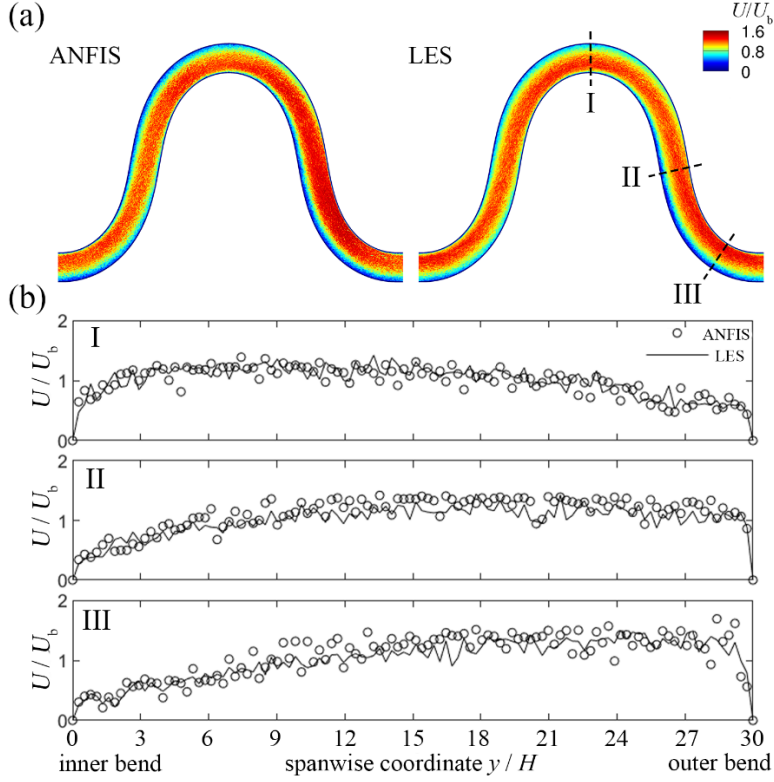

**Figure 20:** Instantaneous LES results and ANFIS predictions for the 3D flow field of the test case 1 at time $t_{68}$ using the LES-ANFIS hybrid method. (a) shows the contours of velocity magnitude ($U / U_b$) at the free surface of Channel 1 from the top view. In (a), flow is from left to right. (b) shows the profiles of the velocity magnitude in the spanwise direction along the three dashed lines of I, II, and III in (a). In (b), solid lines represent the LES results, while circles represent the predictions of ANFIS.

## 7  Conclusion

The capabilities of an ANFIS machine-learning algorithm to predict the 3D velocity field of four large-scale virtual rivers were examined. The ANFIS was first trained using the LES results of one of the large-scale rivers, i.e., Channel 1. Then, the accuracy of the trained ANFIS was evaluated by comparing 3D flow field realizations with the LES results of all four virtual rivers. Our study shows that the ANFIS can accurately predict the 3D flow field of the four rivers in a single time-step prediction. However, the performance of the ANFIS was reduced when it was used for multiple time-step predictions. To overcome the shortcomings of the ANFIS in multiple time-step predictions, we introduced a new approach that involves the recurrent application of LES and ANFIS. Although more expensive than the ANFIS predictions, the proposed recurrent method is about six times less computationally expensive than the LES, and its predictions for instantaneous 3D realizations of flow field are highly accurate.

The results of this study demonstrate the great potential of ANFIS to learn the underlying physics for conducting affordable and yet reliable predictions to generate 3D realizations of turbulent flood flow in large-scale rivers. Specifically, trained using a small batch of LES-computed flow field data, the ANFIS model could successfully predict flow fields in different flow conditions and river geometries. Moreover, the computational cost of the ANFIS predictions is significantly less than that of numerically solving the Navier-Stokes equations with the LES method. Finally, in a future study, we will apply the ANFIS machine-learning algorithm to predict the time-averaged flow field and second-order turbulent statistics of the large-scale rivers under flood conditions.

**Acknowledgments**

This work was supported by the National Science Foundation (grants EAR-0120914 and EAR-1823530). Computational resources were provided by the College of Engineering and Applied Science at Stony Brook University.

**Data availability statement**

For additional information regarding the data generated in this study, the reader can contact the corresponding author via email: ali.khosronejad@stonybrook.edu.